\documentclass[a4paper]{article}

\usepackage{IBERSPEECH2020}
\usepackage{graphicx}
\usepackage{enumitem}
\usepackage{booktabs}

\usepackage{soul}
\usepackage[dvipsnames]{xcolor}

\graphicspath{ {images/}}
\title{Speech Enhancement for Wake-Up-Word detection in Voice Assistants}
\name{David Bonet$^{1,4}$, Guillermo C\'ambara$^{2,4}$, Fernando L\'opez$^3$, \\ Pablo G\'omez$^3$, Carlos Segura$^4$, Jordi Luque$^4$}
\address{
  $^1$Universitat Politècnica de Catalunya (UPC), Spain\\
  $^2$Universitat Pompeu Fabra (UPF), Spain\\
  $^3$Telefónica, Spain\\
  $^4$Telefónica Research, Spain
  }
  
\email{jordi.luque@telefonica.com }

\begin{document}

\maketitle

\begin{abstract}
Keyword spotting and in particular Wake-Up-Word (WUW) detection is a very important task for voice assistants.
A very common issue of voice assistants is that they get easily activated by background noise like music, TV or background speech that accidentally triggers the device.
In this paper, we propose a Speech Enhancement (SE) model adapted to the task of WUW detection that aims at increasing the recognition rate and reducing the false alarms in the presence of these types of noises. The SE model is a fully-convolutional denoising auto-encoder at waveform level and is trained using a log-Mel Spectrogram and waveform reconstruction losses together with the BCE loss of a simple WUW classification network. A new database has been purposely prepared for the task of recognizing the WUW in challenging conditions containing negative samples that are very phonetically similar to the keyword.
The database is extended with public databases and an exhaustive data augmentation to simulate different noises and environments.
The results obtained by concatenating the SE with a simple and state-of-the-art WUW detectors show that the SE does not have a negative impact on the recognition rate in quiet environments while increasing the performance in the presence of noise, especially when the SE and WUW detector are trained jointly end-to-end.

\end{abstract}

\section{Introduction}

Voice interaction with devices is becoming ubiquitous. Most of them use a mechanism to avoid the excessive usage of resources, a trigger word detector. This ensures the efficient use of resources, using a Speech-To-Text tool only when needed and with the consequent start of a conversation. It is key to only start this conversation when the user is addressing the device, otherwise the user experience is notoriously degraded. Thus, the wake-up-word detection system must be robust enough to avoid wake-ups with TV, music, speech and sounds that do not contain the key phrase.  

A common approach to reduce the impact of this type of noise in the system is the adoption of speech enhancement algorithms. Speech enhancement consists of the task of improving the perceptual intelligibility and quality of speech by removing background noise \cite{loizou2013speech}. 
Its main application is in the field of mobile and internet communications \cite{reddy2019scalable} and related to hearing aids \cite{yang2005spectral}, but SE has also been applied successfully to automatic speech recognition systems \cite{zorilua2019investigation,maas2012recurrent,weninger2015speech}.

Traditional SE methods involved a characterization step of the noise spectrum which is then used to try reduce the noise from the regenerated speech signal. Examples of these approaches are spectral subtraction \cite{yang2005spectral}, Wiener filtering \cite{meyer1997multi} and subspace algorithms \cite{ephraim1995signal}. One of the main drawbacks of the classical approaches is that they are not very robust against non-stationary noises or other type of noises that can mask speech, like background speech. 
In the last years, Deep Learning approaches have been widely applied to SE at the waveform level \cite{rethage2018wavenet,phan2020improving} and spectral level \cite{weninger2015speech,park2016fully}. In the first case, a common architecture falls within the encoder-decoder paradigm. In \cite{pascual2017segan}, authors proposed a fully convolutional generative adversarial network architecture structured as an auto-encoder with U-Net like skip-connections. Other recent work \cite{defossez2020real} proposes a similar architecture at the waveform level that includes a LSTM between the encoder and the decoder and it is trained directly with a regression loss combined with a spectrogram domain loss.

Inspired by these recent models, we propose a similar SE auto-encoder architecture in the time domain that is optimized not only by minimizing waveform and Mel-spectrogram regression losses, but also includes a  task-dependent classification loss provided by a simple WUW classifier acting as a Quality-Net \cite{fu2019learning}. This last term serves as a task-dependent objective quality measure that trains the model to enhance important speech features that might be degraded otherwise.




\section{Speech Enhancement}

Speech enhancement is interesting for triggering phrase detection since it tries to remove noise that could trigger the device, and at the same time improves speech quality and intelligibility for a better detection. In this case, we try to tackle the most common noisy environments where voice assistants are used: TV, music, background conversations, office noise and living room noise. Some of these types of background noise, such as TV and background conversations, are the most likely to trigger the voice assistant and are also the most challenging to remove.

\begin{figure*}[t]
  \centering
  \includegraphics[width=\textwidth]{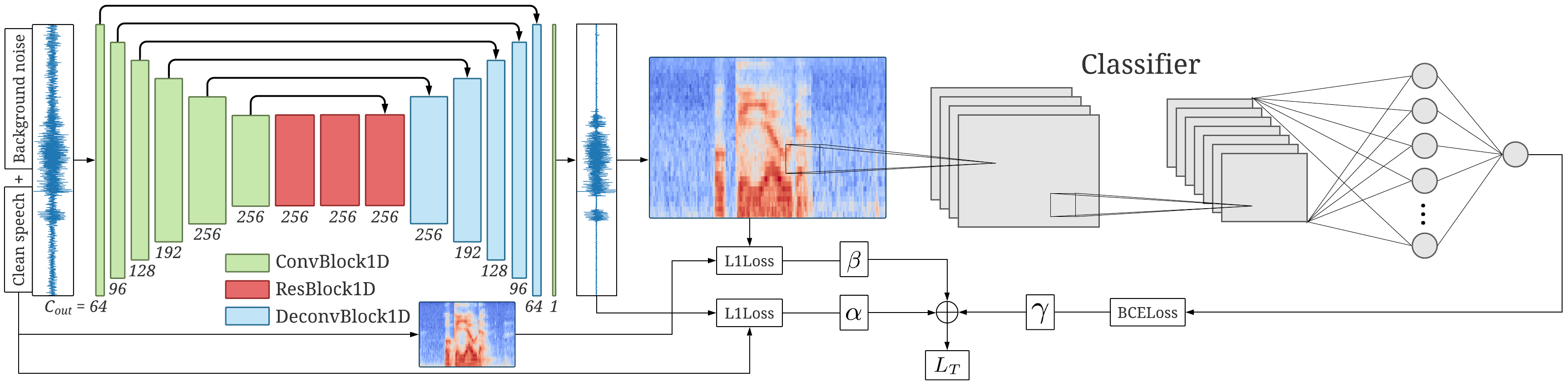}
  \caption{End-to-end SE model at waveform level concatenated with a classifier. Log-Mel Spectrogram and waveform reconstruction losses of the SE model can be used together with the BCE loss of the classifier as a linear combination to train the model.}
  \label{fig:modelo}
\end{figure*}

\subsection{Model} \label{se_model}

Our model has a fully-convolutional denoising auto-encoder architecture with skip connections (Fig. \ref{fig:modelo}), which has proven to be very effective in SE tasks \cite{pascual2017segan}, working end-to-end at waveform level. In training, we input a noisy audio $\boldsymbol{x} \in \mathbb{R}^T$, comprised of clean speech signal $\boldsymbol{y} \in \mathbb{R}^T$ and background noise $\boldsymbol{n} \in \mathbb{R}^T$ so that $\boldsymbol{x}=\lambda \boldsymbol{y}+(1-\lambda)\boldsymbol{n}$.

The encoder compresses the input signal and expands the number of channels. It is composed of six convolutional blocks (ConvBlock1D), each consisting of a convolutional layer, followed by an instance normalization and a rectified linear unit (ReLU). Kernel size $K=4$ and stride $S=2$ are used, except in the first layer where $K=7$ and $S=1$. The compressed signal goes through an intermediate stage where the shape is preserved, consisting of three residual blocks (ResBlock1D), each formed by two ConvBlock1D with $K=3$ and $S=1$ where a skip connection is added from the input of the residual block to its output. The last stage of the SE model is the decoder, where the original shape of the raw audio is recovered at the output. Its architecture follows the inverse structure of the encoder, where deconvolutional blocks (DeconvBlock1D) replace the convolutional layers of the ConvBlock1D with transposed convolutional layers. Skip connections from the encoder blocks to the decoder blocks are also used to ensure low-level detail when reconstructing the waveform.



We use a regression loss function (L1 loss) at raw waveform level together with another L1 loss over the log-Mel Spectrogram as proposed in \cite{yamamoto2020parallel} to reconstruct a "cleaned" signal $\hat{\boldsymbol{y}}$ at the output. Finally, we include the classification loss (BCE Loss) when training the SE model jointly with the classifier or concatenating a pretrained classifier at its output. Thus, we also try to optimize the SE model to the specific task of WUW classification. Our final loss function is defined as a linear combination of the three losses:
\begin{equation} \label{eq:1}
    L_{T}=\alpha L_{raw}(\boldsymbol{y},\hat{\boldsymbol{y}})+\beta L_{spec}(S(\boldsymbol{y}),S(\hat{\boldsymbol{y}}))+\gamma L_{BCE}    
\end{equation}
where $\alpha$, $\beta$ and $\gamma$ are hyperparamters weighting each loss term and $S(\cdot)$ denotes the log-Mel Spectrogram of the signal, which is computed using 512 FFT bins, a window of 20 ms with 10 ms of shift and 40 filters in the Mel Scale.

\section{Methodology}

\subsection{Databases}

The database used for conducting the experiments here presented consists of WUW samples labeled as positive, and other non-WUW samples labeled as negative. Since the chosen keyword is "OK Aura", which triggers Telefónica's home assistant, Aura, positive samples are drawn from company's in-house databases. Some of the negative samples have been also recorded in such databases, but we also add speech and other types of acoustic events from external data sets, so the models gain robustness with further data. Information about all data used is detailed in this section.

\subsubsection{OK Aura Database}


In a first round, around 4300 WUW samples from 360 speakers have been recorded, resulting in 2.8 hours of audio. 
Furthermore, office ambient noise has been recorded as well, with the aim of having samples for noise data augmentation. The second data collection round has been done in order to study and improve some sensitive cases where WUW modules typically underperform. For instance, such dataset contains rich metadata about positive and negative utterances, like room distance, speech accent, emotion, age or gender. Furthermore, the negative utterances contain phonetically similar words to "OK Aura", since these are the most ambiguous to recognize for a classifier. 
Detailed information about data acquisition is explained in the following subsection.

\subsubsection{Data acquisition}

A web-based Jotform form has been designed for data collection. Such form is open and actually still receiving input data from volunteers, so readers are also invited to contribute to the dataset\footnote{https://form.jotform.com/201694606537056}. 
Until the date of this work, 1096 samples from 80 speakers have been recorded, which consists of 1.2 hours of audio. Volunteers are asked to pronounce various scripted utterances at a close distance and also at two meters from the device mic. 
The similarity levels are the following:

\begin{enumerate}[leftmargin=15px]
\itemsep0em 
  \item Exact WUW, in an isolated manner: \textit{OK Aura.}
  \item Exact WUW, in a context: \textit{Perfecto, voy a mirar qué dan hoy. OK Aura.}
  \item Contains "Aura": \textit{Hay un aura de paz y tranquilidad.}
  \item Contains "OK": \textit{OK, a ver qué ponen en la tele.}
  \item Contains similar word units to "Aura": \textit{Hola Laura.}
  \item Contains similar word units to "OK": \textit{Prefiero el hockey al baloncesto.}
  \item Contains similar word units to "OK Aura": \textit{Porque Laura, ¿qué te pareció la película?}
\end{enumerate}


\subsubsection{External data}

General negative examples have been randomly chosen from the publicly available Spanish Common Voice corpus \cite{ardila2019common} that currently holds over 300 hours of validated audio. However, we keep a 10:1 ratio between negative and positive samples, since such ratio proves to yield good results in \cite{hou2020mining}, thus avoiding bigger ratios that lead to increasing computational times. Therefore, we have used a Common Voice partition consisting of 55h for training, 7h for development and 7h for testing.

Background noises were selected from various public datasets according to different use case scenarios. Living room background noise (HOME-LIVINGB) from the QUT-NOISE Database \cite{dean2010qut}, TV audios from the IberSpeech-RTVE Challenge \cite{lleida2019albayzin}, and music\footnote{https://freemusicarchive.org/} and conversations\footnote{http://www.podcastsinspanish.org/} from free libraries.

\subsubsection{Data processing}

All the audio samples are monoaural signals stored in Waveform Audio File Format (WAV) with a sampling rate of 16kHz. The speech data that has been collected was processed with a Speech Activity Detection (SAD) module producing timestamps where speech occurs. For this purpose the tool from pyannote.audio \cite{Bredin2020} has been used, which has been trained with the AMI corpus \cite{carletta2007unleashing}. This helped us to only use the valid speech segments of the audios we collected.

As features to train the models we mainly used two: Mel-Frequency Cesptral Coefficients (MFCCs) and log-Mel Spectrogram. The MFCCs were constructed first filtering the audio with a band pass filter (20Hz to 8kHz) and then, extracting the first thirteen coefficients with 100 milliseconds of windows size and frame shifting of 50 milliseconds. The procedure to extract the log-Mel Spectrogram ($S(\cdot)$) is detailed in section \ref{se_model}.

Train, development and test partitions are split ensuring that neither speaker nor background noise is repeated between partitions, trying to maintain a 80-10-10 proportion, respectively. Total data, containing internal and external datasets, consists of 50.737 non-WUW samples and 4.651 WUW samples.

\subsection{Data augmentation} \label{augmentation}

Several Room Impulse Responses (RIR) were created based on the Image Source Method (ISM) \cite{allen1979image}, for a room of dimensions $(L_x,L_y,L_z)$ where $2 \leq L_x \leq 4.5, 2 \leq L_y \leq 5.5, 2.5 \leq L_z \leq 4$ meters, with microphone and source randomly located at any $(x,y)$ point within a height of $0.5 \leq z \leq 2$ meters. Every TV and music original recordings were convolved with different RIRs to simulate the signal picked up by the microphone of the device in the room.

Adding background noise to clean speech signals is the main data augmentation technique used in the training stage. We use background noises of different scenarios (TV, music, background conversations, office noise and living room noise) and a wide range of SNRs to improve the performance of the models against noisy environments. In each epoch, we create different noisy samples by randomly selecting a sample of background noise for each speech event and combining them with a randomly chosen SNR in a specified range.


\subsection{Wake-Up Word Detection Models} \label{wuw_models}

With the aim of assessing the quality of the trained SE models, we use several trigger word detection classifier models, reporting the impact of the SE module at WUW classification performance. The WUW classifiers used here are a LeNet, a well-known standard classifier, easy to optimize  \cite{lecun2015lenet}; \verb|Res15|, \verb|Res15-narrow| and \verb|Res8| based on a reimplementation by Tang and Lin \cite{tang2017honk}  of Sainath and Parada's Convolutional Neural Networks (CNNs) for keyword spotting \cite{sainath2015convolutional}, using residual learning techniques with dilated convolutions \cite{tang2017deep}; a \verb|SGRU| and \verb|SGRU2|, two Recurrent Neural Network (RNNs) models, based on the open source tool named Mycroft Precise\cite{mycroftprecise2019}, which is a lightweight wake-up-word detection tool implemented in TensorFlow. These are two bigger variations that we have implemented in PyTorch. We also use a \verb|CNN-FAT2019|, a CNN architecture adapted from a kernel \cite{fat2019} in Kaggle’s FAT 2019 competition \cite{fonseca2019audio}, which has shown good performance in tasks like audio tagging or detection of gender, identity and speech events from pulse signal \cite{cambara2020detection}.




\begin{figure}[hb]
  \centering
  \includegraphics[width=1.0\linewidth]{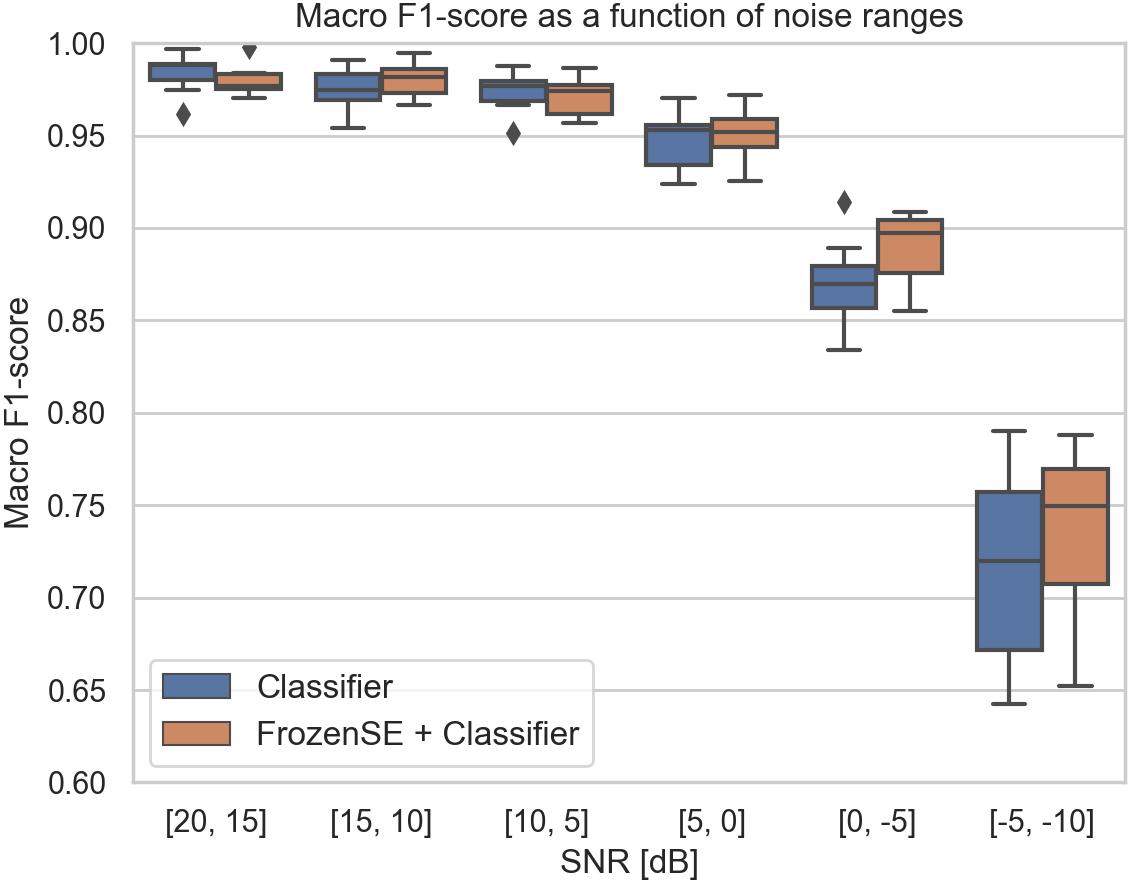}
  \vspace{-15px}
  \caption{Macro F1-score box plot for different SNR ranges. Classifiers trained with low noise ($[5, 30]$ dB SNR).}
  \label{fig:boxplots_1}
  \vspace{-5px}
\end{figure}

\begin{figure}[hb]
  \centering
  \includegraphics[width=1.0\linewidth]{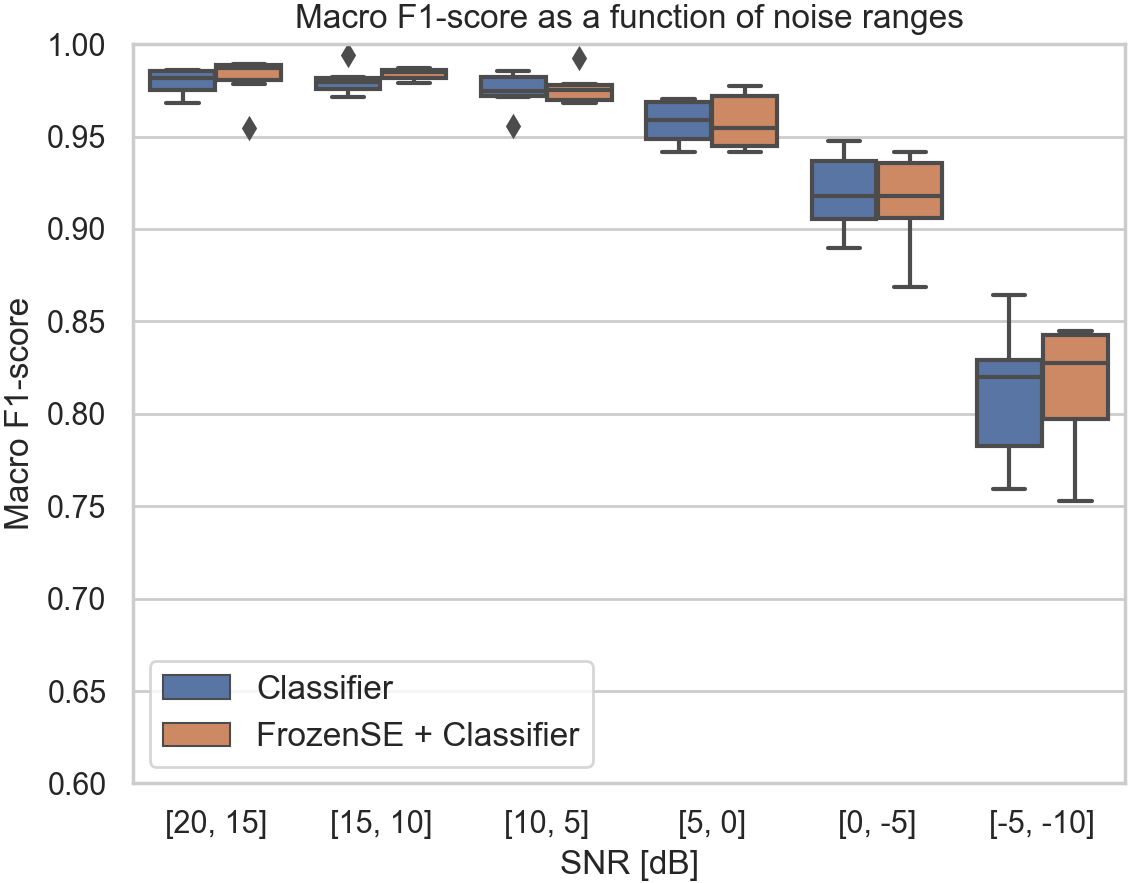}
  \vspace{-15px}
  \caption{Macro F1-score box plot for different SNR ranges. Classifiers trained with a very wide range of noise ($[-10, 50]$ dB SNR).}
  \label{fig:boxplots_2}
  \vspace{-5px}
\end{figure}

\subsection{Training} \label{section:training}

Speech signals and background noises are combined randomly following the procedure explained in \ref{augmentation} with a given SNR range. The SE model is trained to cover a wide SNR range of $[-10,50]$ dBs, whereas WUW models are trained to cover two scenarios: a classifier trained with the same SNR range as the SE model, and a classifier less aware of noise with a narrower SNR range of $[5,30]$ dBs. This way, it is possible to study the impact of the SE model regarding if the classifier has been trained with more or less noise. 


Data imbalance is addressed balancing the classes in each batch using a weighted sampler.  
We use a fixed window length of 1.5 seconds based on the annotated timestamps for our collected database, and random cuts for the rest of the Common Voice samples.

All the models are trained with early stopping based on the validation loss with 10 epochs of patience. We use the Adam optimizer with a learning rate of 0.001 and a batch size of 50. 
Loss (\ref{eq:1}) allows to train the models in multiple ways and we define different SE models and classifiers based on the loss function used:

\begin{enumerate}[label=\alph*),leftmargin=15px]
  \itemsep0em 
    \item \label{caso_a} Classifier: we remove the auto-encoder from the architecture (Fig. \ref{fig:modelo}) and train any of the classifiers using the noisy audio as input: $\alpha=\beta=0$ and $\gamma=1$
    \item \label{caso_b} SE model (SimpleSE): we remove the classifier from the architecture and optimize the auto-encoder based on the reconstruction losses only: $\alpha=\beta=1$ and $\gamma=0$
    \item \label{caso_c} SE model + frozen classifier (FrozenSE): operations of the classifier are dropped from the backward graph for gradient calculation, optimizing only the SE model for a given pretrained classifier (LeNet). $\alpha=\beta=\gamma=1$
    \item \label{caso_d} SE model + classifier (JointSE): auto-encoder and LeNet are trained jointly using the three losses: $\alpha=\beta=\gamma=1$
\end{enumerate}

\subsection{Tests}

All the models take as input windows of 1.5 seconds of audio, to ensure that common WUW utterances are fully within it, since the average "OK Aura" is about 0.8 seconds long. Therefore, we perform an atomic test evaluating if a single window contains the WUW or not. Both negative and positive samples are assigned a background noise sample with which they are combined with a random SNR between certain ranges, as described in section \ref{section:training}.

Given the output scores of the models, the threshold to decide if a sample test is a WUW or not is chosen as the one yielding the biggest difference between true and false positive rates, based on Youden's J statistic \cite{youden1950index}. Once the threshold is decided, macro F1-score is computed in order to balance WUW/non-WUW proportions in the results. We average such scores across all WUW classifiers described in section \ref{wuw_models}, for every SNR range.


\section{Results}



Figure \ref{fig:boxplots_1} illustrates the improvement of the WUW detection in noisy scenarios by concatenating our FrozenSE model with all WUW classifiers described in section \ref{wuw_models} trained with low noise ($[5, 30]$ dB SNR), which we could find in simple voice assistant systems. Applying SE in quiet scenarios maintains fairly good results, and improves them in lower SNR ranges.

If we train the classifiers with more data augmentation ($[-10, 50]$ dB SNR), the baseline classifier results for noisier scenarios improve. Results using the FrozenSE do not decrease but the improvement in ranges of severe noise is not as large as in Figure \ref{fig:boxplots_1}, see Figure \ref{fig:boxplots_2}.

\begin{figure}[t]
  \centering
  \includegraphics[width=1.0\linewidth]{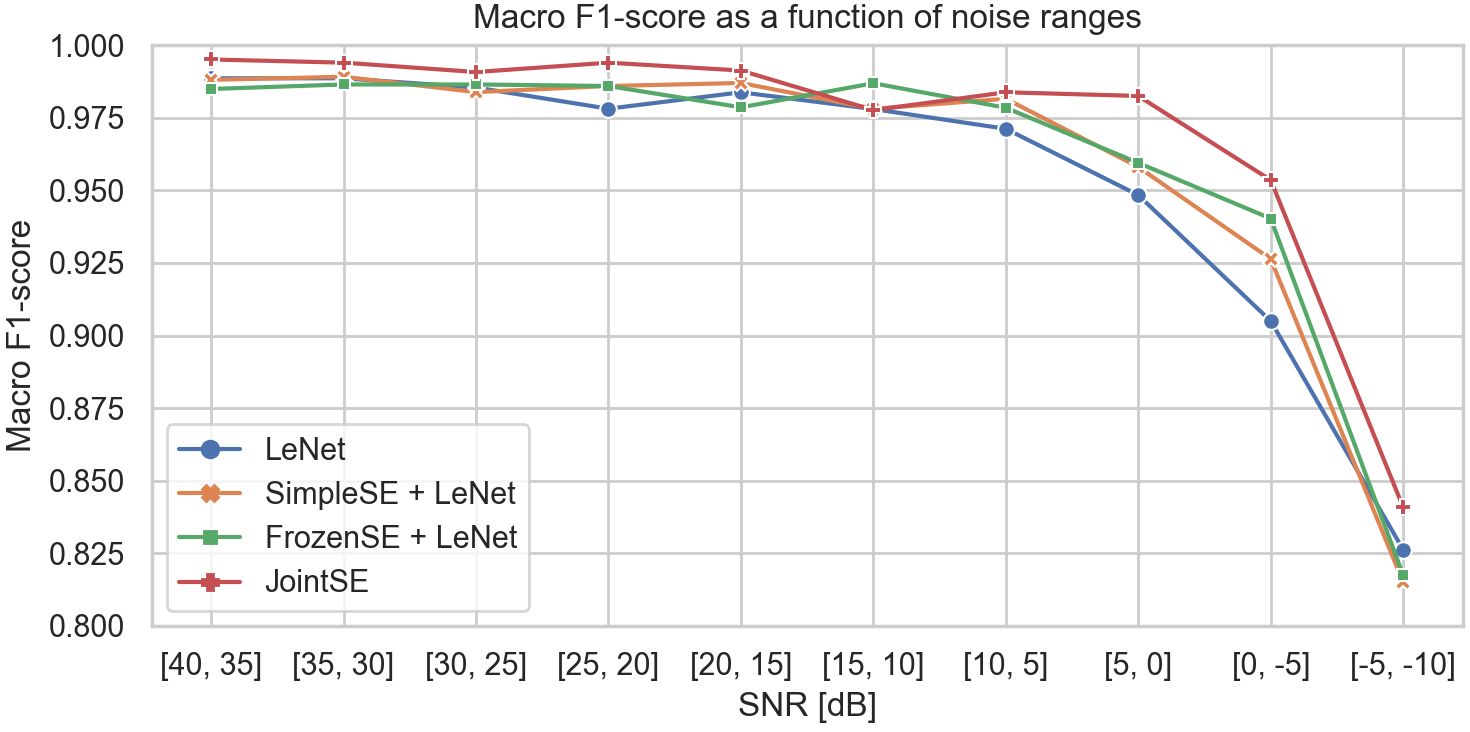}
  \vspace{-15px}
  \caption{Comparison of different training methods for the SE models and LeNet classifier, in terms of the macro F1-Score for different SNR ranges. All models trained in the range of $[-10, 50] dB$ SNR.}
  \label{fig:curvas}
  \vspace{-15px}
\end{figure}

In section \ref{section:training} we have defined the parameters of the loss function (\ref{eq:1}) to train a classifier (case \ref{caso_a}), and different approaches to train the SE model, either standalone (\ref{caso_b}, \ref{caso_c}) or in conjunction with the classifier (\ref{caso_d}). In Figure \ref{fig:curvas} we can see how JointSE performs better than all the other cases in almost every SNR range. From 40 dB to 10 dB of SNR, the results are very similar for the 4 models. In contrast, in the noisiest ranges we can see how the classifier without SE model is the worst performer, followed by the SimpleSE case where only the waveform and spectral reconstruction losses are used. We found that the FrozenSE case, which includes the classification loss in the training stage, improves the results for the wake-up-word detection task. However, the best results are obtained with the JointSE case where the SE model + LeNet are trained jointly using all three losses. 

We compared the WUW detection results of our JointSE with other SOTA SE models (SEGAN \cite{pascual2017segan} and Denoiser \cite{defossez2020real}), followed by a classifier (data augmented LeNet) in different noise scenarios. In Table \ref{tab:sota}, it can be observed how when training the models together with the task loss, the results in our setup are better than with other more powerful but more general SE models, since there is no mismatch between the SE and classifier in the end-to-end and it is also more adapted to common home noises. JointSE improves the detection over the no SE model case, especially in scenarios with background conversations, loud office noise or loud TV, see Table \ref{tab:noise_type}.

\begin{table}[ht]
\caption{Macro F1-score enhancing the noisy audios with SOTA SE models and using a LeNet as a classifier.}
\label{tab:sota}
\centering 
\resizebox{\linewidth}{!}{%
\begin{tabular}{@{}l@{}l@{\hspace{2mm}}l@{\hspace{3mm}}l@{\hspace{2mm}}l@{\hspace{2mm}}l@{}}    
\toprule
\textbf{SNR $[dB]$} && \textbf{No SE} & \textbf{SEGAN} & \textbf{Denoiser} & \textbf{JointSE} \\
\midrule
$[20, 10]$ & Clean      & $0.980$ & $0.964$ & $0.980$  & \boldsymbol{$0.990$}\\ 
$[10, 0]$ & Noisy       & $0.969$ & $0.940$ & $0.955$  & \boldsymbol{$0.972$}\\  
$[0, -10]$ & Very noisy & $0.869$ & $0.798$ & $0.851$  & \boldsymbol{$0.902$}\\
\bottomrule
 \hline
\end{tabular}}
\vspace{-5px}
\end{table}

\begin{table}[ht]
\vspace{-15px}
\caption{Macro F1-score percentage difference between JointSE and LeNet without any SE module, for different background noise types. Positive values mean that the JointSE score is bigger than the single LeNet's.}
\label{tab:noise_type}
\centering 
\resizebox{\linewidth}{!}{%
\begin{tabular}{@{}l@{}l@{\hspace{2mm}}c@{\hspace{2mm}}c@{\hspace{2mm}}c@{\hspace{3mm}}c@{\hspace{3mm}}c@{}}    
\toprule
\textbf{SNR $[dB]$} && \textbf{Music} & \textbf{TV} & \textbf{Office} & \textbf{Living Room} & \textbf{Conversations} \\
\midrule
$[20, 10]$ & Clean      & $1.0$ & $-0.9$  & $1.4$ & $0.4$ & \boldsymbol{$2.3$}\\ 
$[10, 0]$ & Noisy       & $0.0$ & $-1.2$  & $0.8$ & $0.4$ & \boldsymbol{$1.9$}\\  
$[0, -10]$ & Very noisy & $0.5$ & $3.9$  & \boldsymbol{$11.2$} & $3.1$ & $3.8$\\
\bottomrule
 \hline
\end{tabular}}
\vspace{-5px}
\end{table}






\section{Conclusions}

In this paper we proposed a SE model adapted to the task of WUW in voice assistants for the home environment. The SE model is a fully-convolutional denoising auto-encoder at waveform level and it is trained using a log-Mel Spectrogram and waveform regression losses together with a task-dependent WUW classification loss. Results show that for clean and slightly noisy conditions, SE in general does not bring a substantial improvement over a classifier trained with proper data augmentation, 
but in the case of very noisy conditions SE does improve the performance, especially when the SE and WUW detector are trained jointly end-to-end.










\bibliographystyle{IEEEtran}
\bibliography{mybib}

\end{document}